\documentclass{article}

\usepackage{jheppub}

\usepackage{amsmath}
\usepackage{booktabs}
\usepackage{dcolumn}
\usepackage{graphicx}
\usepackage[utf8]{inputenc}
\usepackage{siunitx}
\usepackage{xcolor}
\usepackage{xspace}

\newcommand{\GeV}{\ensuremath{\mathrm{GeV}\xspace}}
\newcommand{\MeV}{\ensuremath{\mathrm{MeV}\xspace}}
\newcommand{\tildeVub}{\ensuremath\tilde{V}_{ub}}
\newcommand{\wc}[2][\ensuremath{\ell}]{\ensuremath{\mathcal{C}_{#2}^{#1}}}
\newcommand{\op}[2][\ensuremath{\ell}]{\ensuremath{\mathcal{O}_{#2}^{#1}}}

\newcommand{\eg}{\textit{e.g.}\xspace}
\newcommand{\ie}{\textit{i.e.}\xspace}
\newcommand{\vecx}{\ensuremath{\vec{x}}\xspace}
\newcommand{\vecth}{\ensuremath{\vec{\vartheta}}\xspace}
\newcommand{\vecnu}{\ensuremath{\vec{\nu}}\xspace}
\newcommand{\dynesty}{\texttt{dynesty}\xspace}
\newcommand{\EOS}{\texttt{EOS}\xspace}

\newcommand{\sklearn}{\texttt{scikit-learn}\xspace}

\makeatletter
\let\oldtheequation\theequation
\renewcommand\tagform@[1]{\maketag@@@{\ignorespaces#1\unskip\@@italiccorr}}
\renewcommand\theequation{(\oldtheequation)}
\makeatother

\begin{document}

\title{Toward a complete description of $\boldsymbol{b \to u \ell^- \bar\nu}$ decays within the Weak Effective Theory}
\author[a]{Domagoj Leljak,}
\emailAdd{domagojleljak@gmail.com}
\author[a]{Bla\v zenka  Meli\'c,}
\emailAdd{melic@irb.hr}
\author[b]{Filip Novak,}
\emailAdd{filip.novak@tum.de}
\author[c]{M\'eril Reboud,}
\emailAdd{merilreboud@gmail.com}
\author[c]{Danny van Dyk}
\emailAdd{danny.van.dyk@gmail.com}

\affiliation[a]{Rudjer Boskovic Institute, Division of Theoretical Physics, Bijeni\v cka 54, HR-10000 Zagreb, Croatia}
\affiliation[b]{Physik Department T31, Technische Universit\"at M\"unchen, 85748 Garching, Germany}
\affiliation[c]{Institute for Particle Physics Phenomenology and Department of Physics, Durham University, Durham DH1 3LE, UK}

\abstract{
We fit the available data on exclusive semileptonic $b\to u\ell^-\bar\nu$ decays within
the Standard Model and in the Weak Effective Theory.
Assuming Standard Model dynamics, we find $|V_{ub}| = 3.59^{+0.13}_{-0.12} \times 10^{-3}$.
Lifting this assumption, we obtain stringent constraints
on the coefficients of the $ub\ell\nu$ sector of the Weak Effective Theory.
Performing a Bayesian model comparison, we
find that a beyond the Standard Model interpretation is favoured over a Standard
Model interpretation of the available data.
We provide a Gaussian mixture model that enables the efficient use of our fit results
in subsequent analyses beyond the Standard Model, within and beyond the framework
of the Standard Model Effective Field Theory.
}

\begin{flushright}
    EOS-2023-01\\
    IPPP/23/01\\
    TUM-HEP-1452/23\\
    RBI-ThPhys-2023-3
\end{flushright}
\vspace*{-3\baselineskip}

\maketitle

\section{Introduction}

The determinations of the Cabibbo-Kobayashi-Maskawa matrix element
$V_{ub}$ have long shown an intriguing puzzle: past determinations from
inclusive decay $\bar{B}\to X_u\ell^-\bar\nu$ at the $b$ factory experiments
have delivered a markedly different result than determinations from
exclusive decays~\cite[Ch.~12.2.6]{ParticleDataGroup:2022pth}.
A more recent analysis of $\bar{B}\to \pi\ell^-\bar\nu$ decays~\cite{Leljak:2021vte}
-- which are presently the most accurate source for an exclusive determination of $|V_{ub}|$ --
as well as a reanalysis of hadronically-tagged $\bar{B}\to X_u\ell^-\bar\nu$ decays by the Belle Experiment~\cite{Belle:2021eni},
have reduced this tension. Nevertheless, other exclusive determinations from
$\bar{B}\to \lbrace \rho,\omega\rbrace\ell^-\bar\nu$ still show deviations.
Effects beyond the Standard Model (BSM) of particle physics have been considered
as a possible explanation for the observed deviations; see \eg Refs.~\cite{Bernlochner:2014ova,Enomoto:2014cta,Sahoo:2017bdx,Banelli:2018fnx,Colangelo:2019axi,Fleischer:2021yjo,Biswas:2022yvh}.
Such effects can be described without using
any UV-complete model within the framework of the Weak Effective Theory
(WET)~\cite{Aebischer:2017gaw,Jenkins:2017jig,Jenkins:2017dyc}.
Here, we study the constraints on the $ub\ell\nu$ sector of the WET, \ie,
the set of operators with $ub\ell\nu$ flavour quantum numbers that is closed
under the remnant of the Standard Model (SM) gauge group after electroweak symmetry
breaking~\cite{Aebischer:2017ugx}.
The WET Hamiltonian for this $ub\ell\nu$ sector -- with left-handed neutrinos only --
reads
\begin{equation}
    \mathcal{H}^{ub\ell\nu}
        = -\frac{4 G_F}{\sqrt{2}} \tildeVub \sum_i \wc{i} \op{i} + \dots + \text{h.c.}\,.
\end{equation}
Here $\wc{i}$ represent a Wilson coefficient, $\op{i}$ represents a local dimension-six
effective field operator, and the dots indicate contributions by operators of mass dimension larger than six.
When assuming only the presence of left-handed neutrinos, the basis of dimension-six operators
consists of five operators. Our choice of operator basis reads
\begin{equation}
\begin{aligned}
    \op{V,L} & = \big[\bar{u} \gamma^\mu P_L  b\big]\, \big[\bar{\ell} \gamma_\mu      P_L \nu\big]\,, &
    \op{V,R} & = \big[\bar{u} \gamma^\mu P_R  b\big]\, \big[\bar{\ell} \gamma_\mu      P_L \nu\big]\,, \\
    \op{S,L} & = \big[\bar{u}            P_L  b\big]\, \big[\bar{\ell}                 P_L \nu\big]\,, &
    \op{S,R} & = \big[\bar{u}            P_R  b\big]\, \big[\bar{\ell}                 P_L \nu\big]\,, \\
    \op{T}   & = \big[\bar{u} \sigma^{\mu\nu} b\big]\, \big[\bar{\ell} \sigma_{\mu\nu} P_L \nu\big]\,.
\end{aligned}
\end{equation}
Within the SM, $\tildeVub$ corresponds to the CKM matrix element $V_{ub}$, all Wilson coefficients except
$\wc{V,L}$ vanish, and
\begin{equation}
    \wc{V,L} = 1 + \frac{\alpha_e}{\pi} \ln\left(\frac{M_Z}{\mu}\right)\,.
\end{equation}
Throughout this work we renormalize the Wilson coefficients at the scale $\mu = 4.2\,\GeV$.
In a generic BSM scenario, the Wilson coefficients encode the effects of BSM physics at energy scales
above the scale of electroweak-symmetry breaking. As a consequence, information on the
$ub\ell\nu$ Wilson coefficients constrains the parameter space of UV-complete BSM models.

In this study, we provide a comprehensive lepton-flavour-universal analysis of semileptonic
$b\to u$ decays with light leptons, \ie, for $\ell=e,\mu$.
We do not attempt to disentangle semielectronic from semimuonic decays due to a lack of
available data. Throughout this analysis we use $m_\ell = 0$.
This simplification can and should be revisited
once precise experimental results become available; see a similar discussion for the case of
$b\to c\ell^-\bar\nu$ processes in Ref.~\cite{Bobeth:2021lya}.
We strive to make our analysis modular and reproducible, to reduce complexity and efforts needed
to update it in the future. To this end, we
provide the necessary files to reproduce our analysis and results based on the theoretical inputs,
experimental measurements, and software settings used.\\

Although fits of the $ub\ell\nu$ WET Wilson coefficients have been published before, our analysis
does feature novel aspects.
Contrary to Refs.~\cite{Bernlochner:2014ova,Enomoto:2014cta,Sahoo:2017bdx,Banelli:2018fnx}, our
analysis covers the full basis of dimension-six operators in the presence of only left-handed neutrinos,
similar to Refs.~\cite{Colangelo:2019axi,Fleischer:2021yjo,Biswas:2022yvh}.
In addition, our analysis covers substantially larger parameter space, which allows for the possibility
of large contributions due to right-handed currents. None of the aforementioned analyses
account for this possibility.
Moreover, in \autoref{sec:results:constraints}, we provide the necessary information to use our statistical
results in future BSM analyses within the SM Effective Field Theory~\cite{Buchmuller:1985jz,Grzadkowski:2010es}
framework, without the need to repeat a low-energy analysis plagued by a large number of hadronic nuisance parameters.
We consider our analysis a path finder for similar analysis
in other sectors of the WET. In this respect, we provide useful information beyond what has been obtained
in the literature.

\section{Analysis Setup}
\label{sec:setup}

We perform a Bayesian analysis of the available experimental data and theoretical information.
The central object of this analysis is the posterior probability density function (PDF) of our parameters
$\vecx \equiv (\vecth, \vecnu)$, which are comprised of the parameters of interest \vecth
and nuisance parameters \vecnu that arise in the description of hadronic matrix elements relevant
to the measured observables. Bayes' theorem defines the posterior PDF $P(\vecx \, | \, D, M)$:
\begin{equation}
    P(\vecx \, | \, D, M) = \frac{P(D \, | \, \vecx, M) \, P_0(\vecx \, | \, M)}{P(D \, | \, M)}\,.
\end{equation}
Here $D$ represents the data used in the analysis, $M$ represents a fit model,
$P_0(\vecx \, | \, M)$ is the prior PDF of the parameters within that model,
and $P(D \, | \, \vecx, M)$ is the likelihood function.
The various fit models and their respective prior PDFs as well as the available likelihoods
are discussed below in detail.
The normalisation constant $P(D \, | \, M)$ -- also known as the evidence -- reads
\begin{equation}
    Z \equiv P(D \, | \, M) = \int d\vecx \, P(D \, | \, \vecx, M) \, P_0(\vecx \, | \, M)\,.
\end{equation}
For identical datasets $D$, the ratio of evidences of two fit models $M_1$ and $M_2$
is useful for a model comparison. It gives rise to the Bayes factor
\begin{equation}
    K \equiv \frac{P(D \, | \, M_1)}{P(D \, | \, M_2)}\,,
\end{equation}
which \emph{favours} model $M_1$ if $K > 1$ and \emph{disfavours} it otherwise.
The Bayes factor can be interpreted according to the scale provided by Jeffreys~\cite{Jeffreys:1939xee}, ranging from
the favour toward $M_1$ being ``barely worth mentioning'' ($1 < K \leq \sqrt{10}$) to being ``decisive'' ($K > 100$).

\subsection{Experimental Data}
\label{sec:setup:data}

In the present analysis, we use the available experimental data on the exclusive decay processes $\bar{B}^0\to \pi^+\ell\bar\nu$,
$\bar{B}^-\to \rho^{0}\ell^-\bar\nu$, and $\bar{B}^-\to \omega\ell^-\bar\nu$.
Our analysis is not sensitive to isospin symmetry violation, and data for the isospin partners of these
decays is recast under the assumption of exact isospin symmetry. In the following, we will not refer to
charge-specific decays unless relevant for the discussion, and all branching ratios are to be interpreted
as the CP-average of the corresponding quantities.
The concrete data used are:
\begin{description}
    \item[$\boldsymbol{\bar{B} \to \pi \ell^-\bar\nu}$] We use the world average
    for the $q^2$-binned branching ratio spectrum for these decays as published
    by the HFLAV collaboration~\cite{HFLAV:2019otj}.
    This average is based on individual results obtained by the BaBar~\cite{BaBar:2010efp,BaBar:2012thb}
    and Belle~\cite{Belle:2010hep,Belle:2013hlo} experiments.
    It accounts for systematic correlations
    between the individual experimental results.
    The measurements from both experiments have compatible binning schemes and
    obtaining their averaged distributions does not require further assumptions.
    A total of 13 bins in $q^2$ are provided.

    Note that the $p$ value of the HFLAV average is $0.04$, which is low but still above
    our a-priori $p$-value threshold. We investigate the effect of rescaling the
    HFLAV uncertainties by a factor of $1.14$. This factors corresponds to rescaling the
    uncertainties of the averaged data sets in such a way that the average's $p$ value reaches $32\%$.
    We find that this procedure has no significant impact on our results: we find shifts
    to $|V_{ub}|$ at or below the $0.6\%$ level. In the following, we only provide results obtained
    from using the unmodified HFLAV average.

    \item[$\boldsymbol{\bar{B} \to \rho \ell^-\bar\nu}$] We use the world average
    for the $q^2$-binned branching ratio spectrum as published in Ref.~\cite{Bernlochner:2021rel}.
    This spectrum was obtained from measurements performed by the BaBar~\cite{BaBar:2010efp}
    and Belle~\cite{Belle:2013hlo} experiments.
    The measurements from both experiments have compatible binning schemes and
    obtaining their averaged distributions does not require further assumptions.
    A total of 11 bins in $q^2$ are provided.

    Here we work within the commonly used narrow-width approximation for the $\rho$.
    This approximation should be lifted as soon as experimental data on
    the full kinematic distribution for the four-body decay
    $\bar{B} \to \pi\pi \ell^-\bar\nu$ become
    available, which includes the resonant $\rho\ell^-\bar\nu$ final state.
    A substantial body of preparatory theoretical work on this
    subject has been carried out already~
\cite{Faller:2013dwa,Kang:2013jaa,Hambrock:2015wka,Hambrock:2015aor,Boer:2016iez,Cheng:2017smj,Cheng:2017sfk,Feldmann:2018kqr,Leskovec:2022ubd}.
    A recent study of the four-body decay's integrated branching ratio
    by the Belle collaboration is an encouraging first step~\cite{Belle:2020xgu}.
    However, it is \emph{currently} not useful due to persistent large theory uncertainties
    in the relevant hadronic matrix elements needed to disentangle the $\pi\pi$ partial waves.

    \item[$\boldsymbol{\bar{B}^- \to \omega \ell^-\bar\nu}$] Analogously to the $\bar{B} \to \rho\ell^-\bar\nu$ case,
    we use the world average for the $q^2$-binned branching ratio spectrum as published in Ref.~\cite{Bernlochner:2021rel}.
    This spectrum was obtained from measurements performed by the BaBar~\cite{BaBar:2012dvs} and Belle~\cite{Belle:2013hlo} 
    experiments.
    However, in this case the respective binning schemes used by either experiment are not mutually compatible.
    To produce an averaged binned $q^2$ distribution, the authors of Ref.~\cite{Bernlochner:2021rel} use a fit
    to split the two bins reported by Belle. This method requires additional assumption on the dynamics underlying the differential $q^2$
    distribution and the hadronic matrix elements, which are taken from Ref.~\cite{Bharucha:2015bzk}.
    For more details we refer to the original work~\cite{Bernlochner:2021rel}.
    A total of 5 bins in $q^2$ are provided.
    Due to the overall smaller number of signal candidates in the Belle results ($\sim 160$) compared
    to the BaBar results ($\sim 870$), we conclude that the BaBar bins dominate the average and that any
    effects on our BSM analysis are negligible at the current level of precision.
\end{description}
Note that all three of the above likelihoods depend in some way on the assumptions of a SM-like
kinematic distribution. We rely on the experimental analyses to accurately account for the systematic
uncertainties that arise from this assumption. We consider it highly desirable to provide for
reinterpretation-safe publication of likelihoods arising from future experimental analyses,
e.g., by using the framework of HistFactory PDF templates~\cite{Cranmer:2012sba}.
\\

There is a substantial amount of experimental data that we cannot or do not use for a variety of reasons.
Our desire to include any additional data has to be weighed against the impact on the complexity
of the fit, which is already facing a large number of hadronic nuisance parameters.
\begin{itemize}
    \item We do not use data on the fully leptonic decay $\bar{B}^-\to \mu^- \bar\nu$. It has so far only been
    seen with about $2.8\sigma$ significance~\cite{Belle:2019iji} and therefore does not meet the commonly used
    threshold of $5\sigma$.
    Including this likelihood in our fit would add one further nuisance parameter whilst not providing any
    significant amount of information. We confirm this expectation a-posteriori in \autoref{sec:results:CKM}.

    \item We do not currently use data on $B \to \eta^{(\prime)}\ell^-\bar\nu$ decays available from the $B$ factory
    experiments~\cite{BaBar:2012thb,Belle:2017pzx}, which provides the same type of constraints on $|V_{ub}|$ and
    the WET Wilson coefficients as $\bar{B}\to \pi\ell^-\bar\nu$ decays do.
    Moreover, the available numerical results for the relevant form-factors~\cite{Duplancic:2015zna} do not yet meet
    the same standard as the $\bar{B}\to \pi$ form factor ones; see \autoref{sec:setup:nuisances}.
    Without an update to the results of Ref.~\cite{Duplancic:2015zna}, using this data does not provide sufficient
    additional information with respect to the data we already use.
    Hence, at present, the amount of information gained by including $B \to \eta^{(\prime)}\ell^-\bar\nu$ decays
    does not outweigh the added complexity due to $12 + 12$ further hadronic nuisance parameters.
    
    In addition, the data are limited by the small number of $q^2$ bins and a partial lack of experimental
    correlation information.
    Nevertheless, recent preliminary results shown by the Belle experiment~\cite{Belle:2021hah}
    give confidence that the Belle II experiment will be able to contribute significantly more precise measurements
    of the $q^2$-binned branching ratio spectrum in the foreseeable future.
    Consequently, this motivates renewed theory efforts in determining the needed hadronic form factors
    from QCD-based methods and to update the results of Ref.~\cite{Duplancic:2015zna}.

    \item We do not currently use data on inclusive semileptonic $b$ decays~\cite{Belle:2003vfz,Belle:2005viu,BaBar:2005acy,BaBar:2005sgq,BaBar:2011xxm,BaBar:2016rxh,Belle:2021eni}.
    Our rationale is that these decays suffer from the same assumptions of a SM-like distribution of events
    but do not provide sufficient information to overcome this assumption for the purpose of estimating accurate
    systematic uncertainties due to the lack of reconstruction of the hadronic final state.

    \item We cannot currently use data on $\Lambda_b^0 \to p \ell^-\bar\nu$ decays~\cite{LHCb:2015eia} or
    $\bar{B}_s^0\to K^{(*)+} \ell^-\bar\nu$ decays~\cite{LHCb:2020ist}, either.
    Presently, data on these processes are only available from the LHCb experiment,
    using the respective normalisation modes
    $\Lambda_b^0\to \Lambda_c^+ \ell^-\bar\nu$ and $\bar{B}_s^0\to D_s^+ \ell^-\bar\nu$.
    This choice of normalisation introduces the dependence on a large number of further hadronic nuisance parameters and the
    $cb\ell\bar\nu$ sector of the WET. Taking $\Lambda_b^0 \to p \ell^-\bar\nu$ as an example, we would
    incur further $\sim 20$ parameters for this decay, further $\sim 20$ parameters for the normalisation mode
    $\Lambda_b^0\to \Lambda_c^+ \ell^-\bar\nu$, and further $5$ parameters for the $cb\ell\bar\nu$ Wilson coefficients.
    
    Our analysis would strongly benefit from LHCb expressing their existing measurement as a binned
    PDF of the $q^2$ distribution of either decay.
    While the binned PDF is independent of the CKM matrix elements and therefore not relevant to the use in (global) CKM fits, it is dependent on ratios of the $ub\ell\nu$ Wilson coefficients and provides complementary information for WET analyses
    such as we carry out here. The usefulness of the angular observables in the
    effective four-body mode $\bar{B}_s^0\to K^{*+}(\to K \pi)\ell^-\bar\nu$
    in particular for WET analyses has been highlighted in Ref.~\cite{Feldmann:2015xsa}.
\end{itemize}
We expect to revisit these decisions in future updates to our analysis.

\subsection{Parameters of Interest and their Priors}
\label{sec:setup:models}

Our analysis uses a total of three fit models named SM, CKM, and WET.
These models only differ in terms of the parameters of interest $\vecth$. They can be summarised as follows:
\begin{description}
    \item[\textbf{SM}] This fit model fixes $\tildeVub \, \wc{V,L} = 3.67\times 10^{-3}$.
    All other Wilson coefficients are fixed to zero.
    As a consequence, this model has no parameters of interest whatsoever and
    serves as the \emph{null hypothesis} for our model comparisons.
    The chosen normalisation is compatible with the SM assumptions
    and the results of the most recent global fit of the CKM matrix elements
    by the CKMfitter collaboration~\cite{Charles:2004jd}.

    \item[\textbf{CKM}] This fit model defines a single parameter of interest, which reads:
    $|V_{ub}| \equiv |\tildeVub \wc{V,L}|$. The parameter is allowed to float within the interval
    $[3.0, 4.5]\,\times 10^{-3}$ with a uniform prior PDF.
    All other Wilson coefficients are fixed to zero.

    \item[\textbf{WET}] This fit model fixes $\tildeVub = 3.67\times 10^{-3}$; see the description
    of the SM model for our justification.
    All Wilson coefficients are fitted from data with a uniform, uncorrelated prior PDF. Its support
    is given by
    \begin{equation}
    \begin{aligned}
          0  & \leq \wc{V,L} \leq 1\,,    &
          0  & \leq \wc{V,R} \leq 1.1\,,    \\
          0  & \leq \wc{S,L} \leq 0.7\,, &
       -0.7  & \leq \wc{S,R} \leq 0.3\,,  &
       -0.25 & \leq \wc{T}   \leq 0.25\,.
    \end{aligned}
    \end{equation}
\end{description}

\noindent
Our choice of the WET model's prior requires some additional discussion:
\begin{itemize}
    \item All $b\to u\ell^-\bar\nu$ observables are insensitive to the overall phase
    of the effective Hamiltonian.
    Hence, we adopt a phase convention by choosing both $\tildeVub$ and one of the Wilson coefficients to be
    real valued and positive. For this analysis, we choose $\wc{V,L}$ to be positive.\\
    We treat all Wilson coefficients as real-valued quantities, since information on their phase
    relative to $\tildeVub\,\wc{V,L}$ cannot be unambiguously extracted from CP averaged data.
    
    \item We assume $m_\ell = 0$, which can and should be revisited once more
    accurate data become available. In this limit, the scalar and tensor Wilson coefficients
    no longer interfere with the vector Wilson coefficients; for a thorough discussion of
    the phenomenological implications (in the $cb\ell\nu$ sector of the WET) we refer to Ref.~\cite{Bobeth:2021lya}.
    The lack of interference terms causes a loss of sensitivity towards the relative phase
    between these two subsets of Wilson coefficients.
    As a consequence, the number of local modes of the posterior doubles.
    To simplify our analysis, we have restricted our prior to select only half of the local modes
    by choosing $\wc{S,L} \geq 0$. The full posterior should be reconstructed from our results by means
    of the replacement
    \begin{equation}
        \label{eq:setup:models:WCsymmetry}
        P(\vecx \, | \, D) \rightarrow \frac{1}{2}\bigg[P(\vecx \, | \, D) + P(\vecx' \, | \, D)\bigg]\,.
    \end{equation}
    Here $\vecx'$ is obtained by replacing $\wc{S,L}$, $\wc{S,R}$, and $\wc{T}$ with
    $-\wc{S,L}$, $-\wc{S,R}$, and $-\wc{T}$, respectively.
    This procedure does not affect our model comparison by means of the Bayes factor.
    
    \item The allowed ranges for each Wilson coefficient are very loosely based on an upper bound on the integrated $\bar{B}\to \pi\ell^-\bar\nu$ branching ratio.
    For massless leptons, the branching ratio
    can be approximated as a weighted sum of squares of individual Wilson coefficients. The ranges
    above are chosen such that the maximal allowed prediction of the integrated branching ratio does
    not exceed the HFLAV average~\cite{HFLAV:2019otj} by more than $5\sigma$.
    Our choice is purely motivated to speed up the sampling process. We have checked a-posteriori that our choice does not cut into the overall posterior probability in any significant way.
\end{itemize}

\subsection{Hadronic Nuisance Parameters and their Priors}
\label{sec:setup:nuisances}

Hadronic nuisance parameters arise in the description of the hadronic form factors
relevant to the predictions of the observables discussed in \autoref{sec:setup:data}.
These form factors are scalar-valued functions of the momentum transfer, typically $q^2$, that
emerge from the Lorentz decomposition of the hadronic matrix elements of the partonic local
$\bar{u} \Gamma b$ currents. The form factors describe the mismatch between the respective
partonic and the (exclusive) hadronic transitions. For $\bar{B}\to \pi$ transitions, we use the BCL parametrization
of the hadronic form factors~\cite{Bourrely:2008za}, however, with adaptations for the
scalar and tensor form factors as introduced in Ref.~\cite{Leljak:2021vte}.
For both $\bar{B}\to \rho$ and $\bar{B}\to \omega$ transitions, we use the BSZ parametrization~\cite{Bharucha:2015bzk}.
Both parametrizations include the dominant subthreshold poles and employ a ``simplified
$z$ expansion'' approach.

At our present level of knowledge of the various hadronic form factors, we do not face correlations
between parameters pertaining to different hadronic transitions.
Hence, the prior PDF for the hadronic nuisance parameters factorises into three independent
PDFs, one per hadronic final state. Details for each hadronic transition follow:
\begin{description}
    \item[$\boldsymbol{\bar{B}\to \pi}$] We use the same setup as used in a recent analysis~\cite{Leljak:2021vte}
    of $\bar{B}\to \pi$ form factors based on lattice QCD and light-cone sum rules studies.
    The relevant lattice QCD inputs to that analysis are taken from
    Ref.~\cite{FermilabLattice:2015mwy,FermilabLattice:2015cdh,Flynn:2015mha}.
    The light-cone sum rule results are obtained in Ref.~\cite{Leljak:2021vte} directly.
    The total number of $\bar{B}\to \pi$ nuisance parameters is $12$.

    \item[$\boldsymbol{\bar{B}\to \rho}$] We use the results obtained in Ref.~\cite{Bharucha:2015bzk},
    from a fit to form factor data obtained from light-cone sum rules with $\rho$ light-cone distribution amplitudes.
    An independent set of results, based on light-cone distribution amplitudes of the $B$ meson, are
    available~\cite{Gubernari:2018wyi}, but not used in our analysis. Our reasoning is that the
    authors of Ref.~\cite{Gubernari:2018wyi} caution the use of the $B\to \lbrace \pi, \rho, K\rbrace$
    subset of their form factor results, due to issues they encountered in the determination of the
    duality threshold parameters.
    Presently, no lattice QCD results
    are available, although first steps toward extracting $\bar{B}\to \pi\pi$ form factors from
    lattice QCD simulations have been undertaken~\cite{Leskovec:2022ubd}.
    The total number of $\bar{B}\to \rho$ nuisance parameters is $19$.

    \item[$\boldsymbol{\bar{B}\to \omega}$] We use the results obtained in Ref.~\cite{Bharucha:2015bzk},
    from a fit to form factor data obtained from light-cone sum rules. Presently, no lattice QCD results
    are available.
    The total number of $\bar{B}\to \omega$ nuisance parameters is $19$.
\end{description}
The total number of hadronic nuisance parameters is $50$.

\section{Methods and Results}
\label{sec:results}

\begin{table}[t]
\renewcommand{\arraystretch}{1.2}
\centering
\begin{tabular}{c @{\hskip 3em} ccc @{\hskip 2em} c}
    \toprule
                  & \multicolumn{3}{c}{goodness of fit} &
    \\
    fit model $M$ & $\chi^2$  & d.o.f.  & $p$ value [\%] & $\ln Z(M)$
    \\
    \midrule
    SM            & $44.18$   & $48$    & $63.03$         & $372.5 \pm 0.4$
    \\
    CKM           & $43.75$   & $47$    & $60.78$         & $372.4 \pm 0.4$
    \\
    WET           & $36.13$   & $43$    & $76.17$         & $376.5 \pm 0.4$
    \\
    \bottomrule
\end{tabular}
\renewcommand{\arraystretch}{1.0}
\caption{%
    Goodness-of-fit values for the three main fits conducted as part of this analysis.
    We provide $\chi^2 = -2 \ln P(\text{data}\,|\,\vecx^*)$ at the best-fit point
    $\vecx^*$ next to the $p$ value and the natural logarithm of the evidence $\ln Z$.
    We find that the $p$ values associated with each individual likelihood
    are larger than $42\%$.
}
\label{tab:results:gof}
\end{table}

\begin{figure}[tp]
    \centering
    \includegraphics[height=.3\textheight]{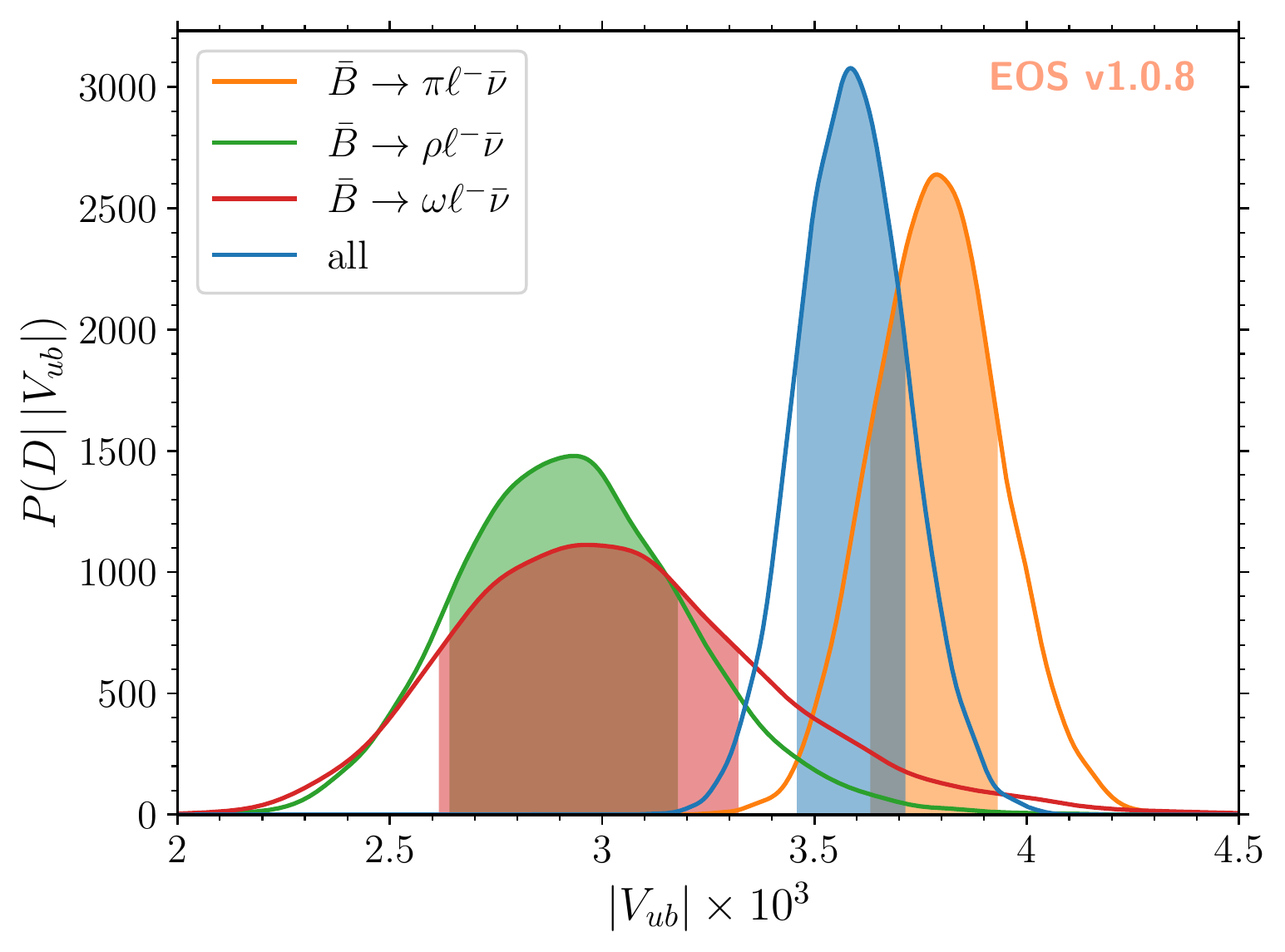}
    \caption{%
        Marginalised one-dimensional posterior densities for $|V_{ub}|$ within the CKM fit model.
        We show our nominal result for the full data set as described in \autoref{sec:setup:data} in blue.
        Additional results for datasets only containing either $\bar{B}\to \pi\ell^-\bar\nu$,
        $\bar{B}\to \rho\ell\bar\nu$, and $\bar{B}\to \omega\ell\bar\nu$ data are shown in orange, green, and red, respectively.
        The shaded areas indicate the central intervals at $68\%$ probability.
    }
    \label{fig:results:CKM:marg-Vub}
\end{figure}
\begin{figure}[tp]
    \centering
    \includegraphics[width=.9\textwidth]{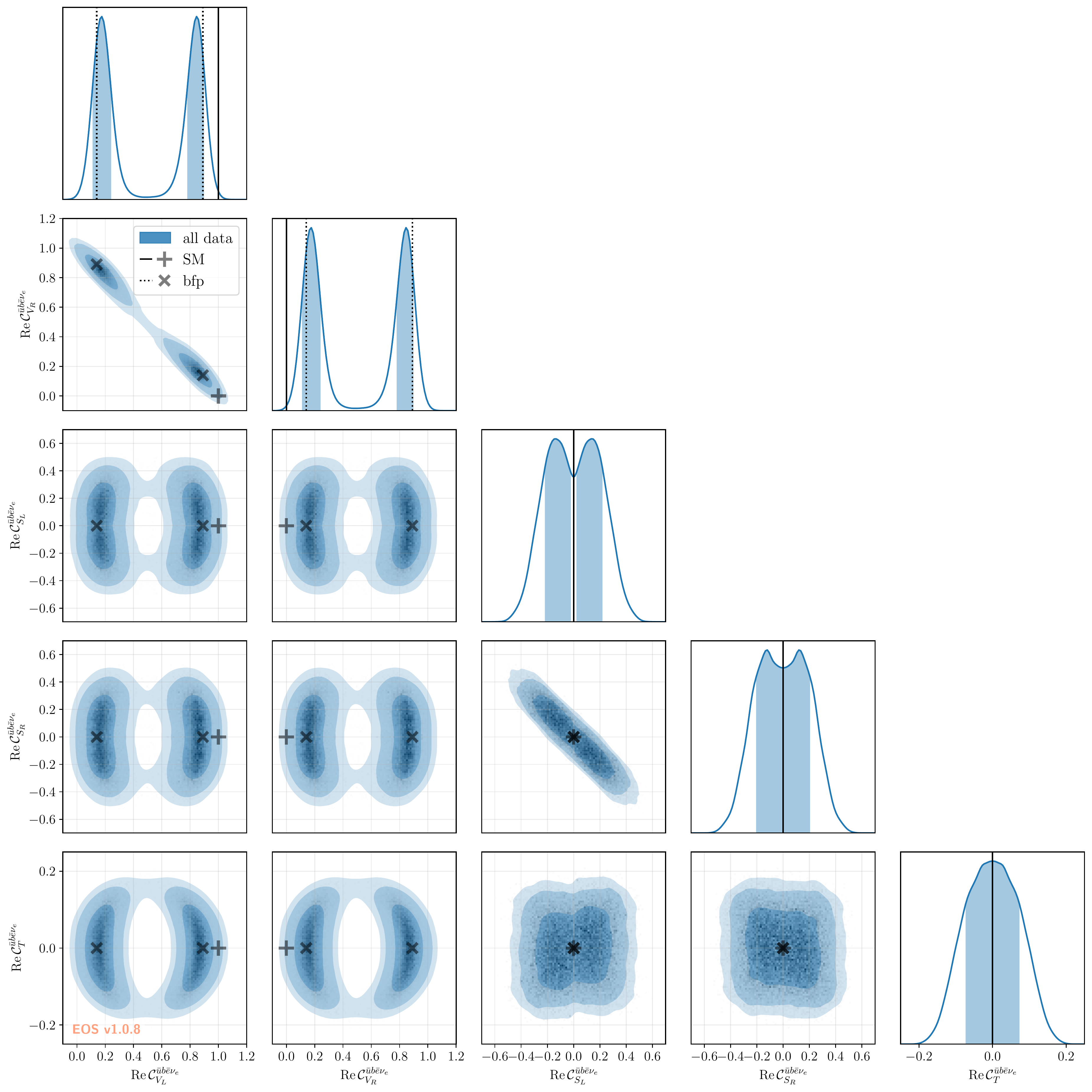}
    \caption{%
        Corner plot of all one-dimensional and two-dimensional marginalised posterior densities within the WET fit model
        for the five parameters of interest.
        We show our nominal result for the full data set as described in \autoref{sec:setup:data}.
        The ``$+$'' and the solid black lines show the SM point $\wc{i} = \delta_{i, (V,L)}$.
        The ``$\times$'' and the dashed black lines show the two best-fit points.
        The blue areas are the $1, 2$ and $3 \sigma$ contours of the posterior distribution obtained from a kernel density estimation.
    }
    \label{fig:results:WET:interest}
\end{figure}

Our overall goals for this analysis are threefold:
\begin{enumerate}
    \item[(a)] to determine whether the available data on exclusive $b\to u\ell\bar\nu$
    processes provides a statistically sound determination of the CKM matrix element
    $|V_{ub}|$ within a global analysis;
    \item[(b)] to determine if a BSM explanation of the data within the WET is favoured or disfavoured
    with respect to the SM explanation;
    \item[(c)] to determine the allowed parameter space of the $ub\ell\nu$ WET sector
    from a global analysis of $b\to u\ell^-\bar\nu$ processes.
\end{enumerate}
To achieve goal (a), we optimise the posterior PDF for each of the three fit models
defined in \autoref{sec:setup:models}. The results are comprised of one best-fit point
$\vecx^*_M$ per model $M$. We provide the global $\chi^2$ values for the three posterior
densities and their corresponding $p$ values in \autoref{tab:results:gof}.
A discussion of $|V_{ub}|$ obtained in our nominal fit is provided in \autoref{sec:results:CKM}.
To achieve goals (b) and (c), we draw samples from each of the three posterior PDFs.
These samples are then used to compute the evidences $Z$ for each posterior PDF,
which are also shown in \autoref{tab:results:gof}. This makes a model comparison possible,
which is carried out in \autoref{sec:results:BSM}.
To achieve goal (c), we marginalise the WET posterior samples onto the
parameter space of interest. We adapt a Gaussian Mixture Model to these samples,
which is available as an ancillary machine readable file~\cite{EOS-DATA-2023-01v2}. The mixture
model enable interested parties to include our results within subsequent BSM analyses
without the need to deal with the full set of hadronic nuisance parameters. This
represents a significant reduction of complexity.
It also ensures that more and more accurate information is
transferred from the low-energy measurements to high-energy model building
analyses than what a Gaussian approximation could accomplish.
The technical steps needed to arrive at the mixture model and further details
are discussed in \autoref{sec:results:constraints}.

All of the above steps are carried out using the \EOS software~\cite{EOSAuthors:2021xpv},
which provides numerical implementations for the theory predictions of observables arising in
$\bar{B}\to \lbrace \pi, \rho, \omega\rbrace \ell^-\bar\nu$ decays, based on the expressions
provided in Ref.~\cite{Duraisamy:2014sna}.
To achieve both accurate and efficient sampling of the posterior densities of more than $50$
parameters,
we have modified \EOS to use dynamical nested sampling~\cite{Higson:2018} as provided
by the \dynesty software~\cite{Speagle:2020,dynesty:v2.0.3}.
This modification is particularly important for the WET fit model, which exhibits a multimodal,
non-Gaussian posterior PDF as shown in \autoref{fig:results:WET:interest}.
Our changes are publicly available as of \EOS version 1.0.8~\cite{EOS:v1.0.8}.

\subsection{CKM Fit}
\label{sec:results:CKM}

We carry out a total of four fits within the CKM fit model.
Our nominal fit uses the combination of all data on $\bar{B}\to \lbrace \pi,\rho,\omega\rbrace\ell^-\bar\nu$
as described in \autoref{sec:setup:data}. We refer to it as the global $|V_{ub}|$ fit.
Three further fits use only experimental data for one of the three exclusive decay modes.
Their purpose is to deepen our understanding of the experimental data and theoretical inputs.
A summary of the goodness-of-fit diagnostics of all four fits are shown in \autoref{tab:results:CKM}.
In all of these fits, the local $p$ values associated to each of the likelihoods are in excess of $42\%$, which ensures that the model correctly reproduces all individual data sets.

We show the one-dimensional marginalised posterior probability densities for the four fits in \autoref{fig:results:CKM:marg-Vub},
in form of Kernel Density Estimates (KDEs).
The posteriors due to the individual decay modes $\bar{B}\to \rho\ell^-\bar\nu$ and $\bar{B}\to \omega\ell^-\bar\nu$
are very broad and their modes are well separated from the mode of the $\bar{B}\to \pi\ell^-\bar\nu$ marginal posterior.
At first glance, this might lead to the conclusion that the $|V_{ub}|$ values extracted from the three individual
decay modes are not mutually compatible.
At face value, averaging these three results (under the assumption of being Gaussian likelihoods)
yields an apparent tension at the $2.3\sigma$ level.
However, we argue that averaging the marginal posterior densities is not the correct procedure,
since it does not account for the possibility of shifting the hadronic nuisance parameters
that are present in the full posterior densities.
Accounting for the nuisance parameters is crucial, since they are fitted to both the experimental
and to the theoretical elements of the posterior. In general, this leads to the nuisance parameters
shifting from their priors' central values.
We therefore hold that the accurate procedure is to first combine the full posterior densities
using the assumption that a single parameter $|V_{ub}|$ can be used to describe all three (uncorrelated) exclusive decay modes.
Indeed, we find that our global fit yields a mutually compatible value for $|V_{ub}| = 3.59^{+0.13}_{-0.12} \times 10^{-3}$
with a $p$ value of $60.78\%$. This value for $|V_{ub}|$ is a central result of our analysis.
It is compatible at the $0.54\sigma$ level with $(3.67^{+0.09}_{-0.07}) \times 10^{-3}$,
which is obtained from a global CKM fit~\cite{Charles:2004jd}.
Our finding that a global fit of $|V_{ub}|$ successfully explains the available data
is corroborated by our global fit in the SM model (see Table 2).
For that model, the $\chi^2$ value deviates from the global $|V_{ub}|$ fit's value by less than one.

Both our individual fit to $\bar{B}\to \pi\ell^-\bar\nu$ and the global $|V_{ub}|$ fit yield
results that are compatible with recent determinations in the presence of dispersive bounds,
with differences well below the $1\sigma$ level~\cite{Martinelli:2022tte}.

Using the posterior samples of the global $|V_{ub}|$ fit, we obtain the following posterior-predictive results
for the three leptonic decay modes
\begin{equation}
\begin{aligned}
    \mathcal{B}(\bar{B}^-\to \tau^-\bar\nu)
        & = \left( 8.28^{+0.61}_{-0.57} \big|_{|V_{ub}|} \pm 0.13 \big|_{f_B} \right) \times 10^{-5}
    \,,\\
    \mathcal{B}(\bar{B}^-\to \mu^-\bar\nu)
        & = \left( 3.72^{+0.27}_{-0.25} \big|_{|V_{ub}|} \pm 0.06 \big|_{f_B} \right) \times 10^{-7}
    \,,\\
    \mathcal{B}(\bar{B}^-\to e^-\bar\nu)
        & = \left( 8.71^{+0.64}_{-0.60} \big|_{|V_{ub}|} \pm 0.14 \big|_{f_B} \right) \times 10^{-12}
    \,.
\end{aligned}
\end{equation}
These results assume the couplings to be lepton-flavour universal,
where we use $f_B = 189.4 \pm 1.4\,\MeV$~\cite{Bazavov:2017lyh}.
Our posterior prediction for $\mathcal{B}(\bar{B}^-\to \mu^-\bar\nu)$ is compatible with the
$90\%$ upper bound set by the Belle experiment~\cite{Belle:2019iji}.
Interpreting this bound as a measurement, Belle obtains
\begin{equation}
    \mathcal{B}(\bar{B}^-\to \mu^-\bar\nu)\big|_{\text{Belle '19}}
        = \left( 5.3 \pm 2.2 \right) \times 10^{-7}
\end{equation}
at $2.8\sigma$ significance.
Our posterior prediction features a small uncertainty, which is smaller than the Belle '19 uncertainty by about a factor of 10.
This finding further supports our decision not to use the Belle data as part of our likelihood at this point.

\begin{table}[t]
    \renewcommand{\arraystretch}{1.2}
    \centering
    \begin{tabular}{c @{\hskip 0em} D{.}{.}{7.2} D{.}{.}{2.0} D{.}{.}{5.1} @{\hskip 0em} c}
        \toprule
                                   & \multicolumn{3}{c}{Goodness of fit} &
        \\
        Data set                   & \multicolumn{1}{c}{$\chi^2$}
                                               & \multicolumn{1}{c}{d.o.f.}
                                                         & \multicolumn{1}{c}{$p$ value [\%]}
                                                                          & \multicolumn{1}{c}{$|V_{ub}|\times 10^3$}
        \\
        \midrule
        $\bar{B}\to \pi\ell\nu$    &  $27.83$  & $31$    & $62.98$         & $3.79^{+0.15}_{-0.15}$
        \\
        $\bar{B}\to \rho\ell\nu$   &  $4.05$   & $10$    & $94.49$         & $2.92^{+0.28}_{-0.25}$ 
        \\
        $\bar{B}\to \omega\ell\nu$ &  $4.20$   &  $4$    & $37.90$         & $3.00^{+0.38}_{-0.32}$
        \\
        all data                   & $43.75$   & $47$    & $60.78$         & $3.59^{+0.13}_{-0.12}$
        \\
        \bottomrule
    \end{tabular}
    \renewcommand{\arraystretch}{1.0}
    \caption{%
        Goodness-of-fit values and results for $|V_{ub}|$ for the global CKM fit (``all data'') and
        three further CKM fits of individual dataset.
        The results for $|V_{ub}|$ correspond to the respective medians of the one-dimensional marginalised
        posterior probability densities and central $68\%$ probability intervals.
    }
    \label{tab:results:CKM}
\end{table}

\subsection{BSM Interpretation}
\label{sec:results:BSM}

We lift our assumption of SM dynamics by fitting the data within the WET model.
We find the posterior to be non-Gaussian and multimodal. It further features almost flat directions that show up as
ring-like contours in the $\wc{V,L}$--\,$\wc{T}$ and $\wc{V,R}$--\,$\wc{T}$ planes, see \autoref{fig:results:WET:interest}.
Moreover, we recover an approximate symmetry under the exchange of $\wc{V,L}$ and $\wc{V,R}$,
as expected from the expressions for the branching ratios in Ref.~\cite{Duraisamy:2014sna}.
We expect that future data on the angular distributions of the three semileptonic decays
used here will help to break this degeneracy of solutions by providing complementary constraints,
due to interference terms $\wc{V,L}\,{\wc{V,R}}^*$.

The WET components of the two main modes of the posterior read
\begin{equation}
\begin{aligned}
    \wc{V,L}  & = 0.14, &
    \wc{V,R}  & = 0.89, \\
    \wc{S,L}  & = 0.00, &
    \wc{S,R}  & = 0.00, &
    \wc{T}    & = 0.00,
\end{aligned}
\end{equation}
and
\begin{equation}
\begin{aligned}
    \wc{V,L}  & = 0.89, &
    \wc{V,R}  & = 0.14, \\
    \wc{S,L}  & = 0.00, &
    \wc{S,R}  & = 0.00, &
    \wc{T}    & = 0.00.
\end{aligned}
\end{equation}
The other modes are obtained by applying the symmetry discussed around \autoref{eq:setup:models:WCsymmetry}.
All of these modes show the same minimal $\chi^2 = 36.13$.
This represents a marked reduction in the minimal $\chi^2$ value when compared to the SM and CKM fits.
The respective reductions are $\Delta \chi^2 = 8.05$ for five degrees of freedom (d.o.f.) and $\Delta \chi^2 = 7.62$ for four d.o.f.
Interpreted as a log-likelihood ratio test and applying Wilks' theorem, we find that a BSM interpretation within the WET
is mildly favoured over the SM and CKM hypotheses with significances of $1.4\sigma$ and $1.6\sigma$, respectively.
The posterior modes are shown in \autoref{fig:results:WET:interest}, next to
the SM point and contours at $68\%$, $95\%$, and $99\%$ probability
for each of the one-dimensional and two-dimensional marginal posterior PDFs.

Beyond a log-likelihood test, we also perform a Bayesian model comparison by means of the Bayes factor. To this end, we draw posterior samples and evaluate the evidence
using dynamical nested sampling; see the earlier discussion.
We find that the posterior samples show a minor breaking of the posterior's
$\wc{V,L}$---$\wc{V,R}$ symmetry. We explicitly check whether the partial evidences
pertaining to each of the two posterior modes are equal in size. This is the case, within
the estimates of the evidence uncertainty.
From the evidences $Z$ given in \autoref{tab:results:gof}, we obtain
\begin{equation}
\begin{aligned}
    \frac{P(\text{all data} \, | \, \text{WET})}{P(\text{all data} \, | \, \text{SM})}
        & = 55\,, &
    \frac{P(\text{all data} \, | \, \text{WET})}{P(\text{all data} \, | \, \text{CKM})}
        & = 60\,.
\end{aligned}
\end{equation}
Our finding shows that the WET model is significantly more efficient in explaining the available data compared to the SM and CKM models.
This can be understood from the discussion in \autoref{sec:results:CKM}: explaining all the data simultaneously requires
the nuisance parameters to deviate from their prior values. Lifting the assumption of SM dynamics reduces
the need to shift the nuisance parameters and allows for larger posterior values on average, thereby
increasing the evidence.
Using Jeffreys' interpretation of the Bayes factor, we find the data to be \emph{strongly in favour} of the BSM
interpretation over either the SM null hypothesis or the CKM hypothesis.

Using the posterior samples of the global $ub\ell\nu$ WET fit, we obtain
posterior-predictive distributions for the leptonic decay modes of the $\bar{B}^-$
meson. In contrast to the results in the CKM fit model, we find very heavy-tailed
distributions. This can be understood, since within the WET description the
scalar operators $\op{S,L}$ and $\op{S,R}$ can contribute without any helicity
suppression. The tails extend well into the region excluded by the Belle
experiment~\cite{Belle:2019iji}. We provide the upper bound at $90\%$ probability
as:
\begin{equation}
\begin{aligned}
    \mathcal{B}(\bar{B}^-\to \tau^-\bar\nu)
        & < 5.67 \times 10^{-4}
    \,,\\
    \mathcal{B}(\bar{B}^-\to \mu^-\bar\nu)
        & < 5.56 \times 10^{-4}
    \,,\\
    \mathcal{B}(\bar{B}^-\to e^-\bar\nu)
        & < 5.53 \times 10^{-4}
    \,.
\end{aligned}
\end{equation}
Using the Belle data on the muonic decay would therefore have a noticeable
impact on our results and likely curtail the WET posterior in the scalar
Wilson coefficients.
Unfortunately, we currently do not see a way how to
faithfully include the data within our experimental likelihood
based on publicly available information.
Indeed, a non-zero BSM contribution would also impact the backgrounds of this analysis
-- currently dominated by $b\to u$ transitions --
and therefore require a complete reinterpretation of the analysis dataset.

\subsection{Constraints on BSM Models}
\label{sec:results:constraints}

To ensure that our results can be used in an accurate and computationally efficient manner within
subsequent studies,
we provide a Gaussian Mixture Model (GMM) of the marginalised five-dimensional posterior of interest within the
WET fit model. Our approach is of general interest to the field, since it avoids re-running a complicated,
computationally expensive statistical analysis. Hence, we describe in some detail our approach
to producing and validating the GMM.

Our notation for the GMM PDF reads:
\begin{equation}
    P_\text{GMM}(\vecth \, | \, \text{data})
        = \sum_{n=1}^{N} \alpha_n \, \mathcal{N}(\vecth \, | \, \vec{\mu}_n, \Sigma_n)\,.
\end{equation}
Here $\vecth = (\wc{V,L}, ..., \wc{T})$ represents the WET Wilson coefficients, $N$ is the total number of Gaussian
components, $\mathcal{N}$ is a multivariate Gaussian probability density, and $\alpha_n$,
$\vec{\mu}_n$, and $\Sigma_n$ are the relative weight, location, and covariance of the $n$th component.

To obtain the parameters $\lbrace \alpha_n, \vec{\mu}_n, \Sigma_n\rbrace$ of the GMM, we use the publicly
available \sklearn software~\cite{scikit-learn}.
The GMM is most conveniently obtained for unweighted posterior samples.
To this end, we resample from the WET posterior density with equal weights using built-in functions of the \dynesty software.

We validate our result for the GMM parameters by computing the normalised Shannon entropy,
\begin{equation}
\begin{aligned}
    \mathcal{P}
        & \equiv \frac{1}{M} \exp\left[-\sum_{m=1}^M \overline{\omega}_m \ln \overline{\omega}_m\right]\,, &
    0 \leq \mathcal{P} \leq 1\,,
\end{aligned}
\end{equation}
also known as the perplexity, that we evaluate on $M=5\times 10^4$ samples generated by the GMM.
The sample weights $\overline{\omega}_m$ iterate over the posterior samples $\lbrace \vartheta_1, \dots, \vartheta_M \rbrace$ and read
\begin{equation}
\begin{aligned}
    \overline{\omega}_m
        & \equiv \frac{\omega_m}{\sum_{m=1}^M \omega_m}\,, &
    \omega_m
        & \equiv \frac{P(\vartheta_m \, | \, \text{all data})}{P_\text{GMM}(\vartheta_m \, | \, \text{all data})}\,.
\end{aligned}
\end{equation}
A perplexity close to unity indicates very good agreement between the density model and the true probability density.
Having no access to the values of the marginalised posterior, we approximate it by means of a KDE,
as already used in \autoref{fig:results:WET:interest}.

We fit the GMM to the posterior samples using a variational Bayes approach
with $N = 20$ components. We obtain a GMM that yields $\mathcal{P} = 0.88$.
Two comments are in order:
\begin{itemize}
    \item We approximate the marginal posterior with a KDE, which is itself a model.
    Hence, our calculation
    of the perplexity compares two models with each other, rather than the posterior with
    one model. As a consequence, we do not expect to be able to reach the optimal perplexity
    of $\mathcal{P} = 1$. The difference between the obtained and the optimal perplexity
    can therefore not fully be attributed to mismodelling by the GMM.
    \item Since there is no test statistics or other statistical diagnosis available
    for this type of modelling, we cannot infer the overall quality of the approximation.
    Based on prior experience with Population Monte Carlo studies~\cite{Kilbinger:2009by,pypmc}, we consider
    a value of $\mathcal{P} = 0.88$ to be good description of the (approximate) marginal posterior.
\end{itemize}
We conclude that, at the current level of accuracy of the available data, the
GMM provides a reasonably and computationally efficient way to transfer information
on the WET Wilson coefficients without having to deal with the large number of hadronic
nuisance parameters.
We therefore provide the GMM parameters as an ancillary YAML file~\cite{EOS-DATA-2023-01v2}.
It is also available within the \EOS software as of version v1.0.8 as the constraint named:
\begin{center}
    \texttt{ublnu::P(WET)\makeatletter@\makeatother{}LMNRvD:2023A}.
\end{center}
The naming of parameters follows the standards set by the Wilson Coefficients eXchange Format~\cite{Aebischer:2017ugx}.
In connection with the \texttt{wilson} software~\cite{Wilson}, the constraint can be used to infer information
on the parameters of the Standard Model Effective Field Theory.
We expect to improve on the software interface in the upcoming v1.1 release of \EOS.
Our results can also be used to constrain the parameters of specific BSM model, as long as these
models do not affect the assumptions underlying the WET, \ie, as long as no new particles and forces
are introduced below the scale of electroweak symmetry breaking.

\section{Summary and Outlook}

We have performed a comprehensive Bayesian analysis of available data on exclusive
$\bar{B}\to \lbrace \pi,\rho,\omega\rbrace \ell^-\bar\nu$ decays within three fit models.
All models fit the data well.

Assuming SM dynamics, we determine the CKM matrix element to be
\begin{equation*}
    |V_{ub}| = (3.59^{+0.13}_{-0.12}) \times 10^{-3}
\end{equation*}
from a global fit to the available data, which is our nominal CKM result.
Further results, arising from more restricted data sets, are briefly discussed in \autoref{sec:results:CKM},
illustrated in \autoref{fig:results:CKM:marg-Vub}, and documented within the ancillary material~\cite{EOS-DATA-2023-01v2}.
We find agreement between the three exclusive decay modes on account
of numerically large uncertainties in the $\bar{B}\to \lbrace \rho, \omega\rbrace$ form factors.

Lifting the assumption of SM dynamics, we perform an analysis within the Weak Effective Theory (WET)
with only left-handed neutrinos. Assuming lepton-flavour universality, we find the data to be better
explained by a non-zero BSM contribution that couples vector-like to $\bar{u}b$.
By means of a Bayesian model comparison, we find that the BSM
interpretation is presently \emph{strongly in favour} of the SM one,
as discussed in \autoref{sec:results:BSM}.

The marginal posterior for the WET parameter space is markedly non-Gaussian, with flat directions
and multiple local modes.
To enable the use of our results without the need to repeat the computationally expensive analysis,
we approximate our results by means of a Gaussian Mixture Model (GMM).
Details on the derivation of this mixture model are discussed in \autoref{sec:results:constraints}.
The model parameters are available in machine-readable form as part of the ancillary material~\cite{EOS-DATA-2023-01v2}.
Future update to our analysis will have the opportunity to improve the quality of the GMM by including further and
complementary constraints on the Wilson coefficients, \eg, from data on the angular distributions.\\

Our analysis can and should be improved in a variety of ways as soon as feasible.
The interpretation of $\bar{B}\to \pi\pi \ell^-\bar\nu$ decays as $\bar{B}\to \rho \ell^-\bar\nu$
decays introduces a hard-to-quantify systematic uncertainty, both on the theory side and
the experimental side. A substantial theory program to address this issues has been put
forward in recent years, and further improvement in this area is now dependent on
experimental data.
As part of this theory program, Refs.~\cite{Hambrock:2015aor,Cheng:2017smj} estimate a universal
change of at maximum $20\%$ to all form factors. Such a universal change would affect the overall scale of
our results for the WET Wilson coefficient; however, it would leave the relative deviation
unchanged. As such, we expect the full treatment of the $\pi\pi$ final state
to affect our results quantitatively, but not qualitatively.
We also look forward to the impact that further information will have. In particular,
we are interested to see the impact of data on the full angular distribution of
$\bar{B}\to [\pi\pi]_\rho\ell^-\bar\nu$ decays as well as
$\Lambda_b \to p\ell^-\bar\nu$ and $\bar{B}_s \to K\pi\ell^-\bar\nu$ decays.
Data on the latter two decays obtained by the LHCb experiment is already part of published
studies, albeit intertwined with data on weak decays from the $cb\ell\nu$ sector of the
WET. We consider revisiting this data and its publication as (binned) PDFs
highly desirable.

\acknowledgments

We thank Aleks Smolkovi\v c and Peter Stangl for numerical comparisons that lead to finding
a bug in the original implementation of $\bar{B}\to \lbrace \rho^0, \omega\rbrace\ell\bar\nu$ decays.
M.R.~thanks Admir Greljo for useful discussions on the paper.
D.L., B.M.~and D.v.D.~acknowledge support from the Alexander von Humboldt Foundation in the
framework of the Research Group Linkage Programme, funded by the German Federal Ministry of
Education. B.M.~has also been supported by the Croatian Science Foundation (HRZZ) project “Heavy
hadron decays and lifetimes” (IP-2019-04-7094).
D.v.D.~acknowledges support by the UK Science and Technology Facilities Council
(grant numbers ST/V003941/1 and ST/X003167/1).

\appendix

\bibliographystyle{JHEP}
\bibliography{b-to-ulnu}

\end{document}